\documentclass[12pt]{article}
\usepackage{epsf}
\def\1ad{\mbox{\normalsize $^1$}}
\def\2ad{\mbox{\normalsize $^2$}}
\def\3ad{\mbox{\normalsize $^3$}}
\def\4ad{\mbox{\normalsize $^4$}}
\def\5ad{\mbox{\normalsize $^5$}}
\def\6ad{\mbox{\normalsize $^6$}}
\def\7ad{\mbox{\normalsize $^7$}}
\def\8ad{\mbox{\normalsize $^8$}}

\def\npb#1#2#3{{ Nucl. Phys.} {\bf B#1} (#2) #3 }
\def\plb#1#2#3{{ Phys. Lett.} {\bf B#1} (#2) #3 }
\def\prd#1#2#3{{ Phys. Rev. } {\bf D#1} (#2) #3 }
\def\prl#1#2#3{{ Phys. Rev. Lett.} {\bf #1} (#2) #3 }

\def\ijmpa#1#2#3{{ Int. J. Mod. Phys.} {\bf A#1} (#2) #3 }

\def\jhep#1#2#3{{ J. High Energy Phys.} {\bf #1} (#2) #3 } 
 
\def\bb#1{{\tt hep-th/#1}}

\def\CA{{\cal A}}

  \def\CW{{\cal W}}

\def\dj{\hbox{d\kern-0.347em \vrule width 0.3em height 1.252ex depth
-1.21ex \kern 0.051em}}

\def\half{{1\over 2}\,}

\def\ket{\rangle}
\def\bra{\langle}

\newcommand{\nn}{\nonumber}
\newcommand{\cf}{\mbox{{\em c.f.~}}}
\newcommand{\ie}{\mbox{{\em i.e.~}}}

\def\shalf{{\mbox{$\half$}}}
\def\ssixth{{\mbox{${1\over 6}$}}}

\def\bra{\langle}
\def\ket{\rangle}
\def\lim{\mbox{{\bf L}} }

\newcommand{\be}{\begin{equation}}
\newcommand{\ee}{\end{equation}}
\newcommand{\ben}{\begin{equation*}}
\newcommand{\een}{\end{equation*}}
\newcommand{\ba}{\begin{eqnarray}}
\newcommand{\ea}{\end{eqnarray}}
\newcommand{\ban}{\begin{eqnarray*}}
\newcommand{\ean}{\end{eqnarray*}}
\newcommand{\brr}{\begin{array}}
\newcommand{\err}{\end{array}}
\newcommand{\bc}{\begin{center}}
\newcommand{\ec}{\end{center}}

\begin{document}

\newcommand{\sheptitle}
{Remarks on Time-Space Noncommutative Field Theories}

\newcommand{\shepauthora}
{{\sc
L. ~Alvarez-Gaum\'e$^{\,a}$, ~ 
 J.L.F.~Barb\'on$^{\,a,}$\,\footnote[1]{On leave from the
Departmento de F\'{\i}sica de Part\'{\i}culas. Universidade de Santiago
de Compostela, Spain.} ~ and ~ R. Zwicky$^{\,a,\,b}$
}}

\newcommand{\shepaddressa}
{\sl
$^{a}$ \,Theory Division, CERN \\
 CH 1211 Geneva 23 \\
 Switzerland 
}

\newcommand{\shepaddressb}
{\sl
$^{b}$ \,Institut f\"ur theoretische Physik. 
Universitaet Z\"urich \\ 
Winterthurerstrasse 190, 
8057 Z\"urich \\
Switzerland 
}

\newcommand{\shepabstract}
 {We propose a physical interpretation of the perturbative breakdown  
of  unitarity in time-like noncommutative field theories
in terms of  production of tachyonic particles. 
These particles may be viewed as a remnant of a continuous spectrum
of undecoupled closed-string modes. In this way,
we give a unified view of
 the string-theoretical and the field-theoretical no-go arguments
against time-like noncommutative theories.  We also perform a quantitative
study of various locality and causality properties 
of noncommutative field theories at the quantum level. 
}

\begin{titlepage}
\begin{flushright}
CERN-TH/2001-052\\
ZH-TH 06/01\\
{\tt hep-th/0103069}\\

\end{flushright}
\vspace{1in}
\begin{center}
{\large{\bf \sheptitle}}
\bigskip\bigskip \\ \shepauthora \\ 
\vspace{0.5in} 
{\it \shepaddressa} \\
\bigskip\bigskip  {\it \shepaddressb} \\
\vspace{0.5in}

{\bf Abstract} \bigskip \end{center} \setcounter{page}{0}
\shepabstract
\vspace{1in}
\begin{flushleft}
CERN-TH/2001-052\\
March 2001 
\end{flushleft}


\end{titlepage}

\subsection*{Introduction}

\noindent

Quantum field theories in noncommutative spaces \cite{conn, class}
  with noncommutative time
coordinate are notoriously ill-defined. Heuristically,  
noncommutative particles of momentum $p^\mu$ 
can be regarded as rigid, extended dipoles 
 oriented along the  four-vector
   $L^{\mu} = \theta^{\mu\nu} p_\nu$,  and 
 interacting through the end-points \cite{dipoles}.
  It is already clear from this
consideration that noncommutative time coordinates imply particles
effectively `extended in time'. Thus, the breakdown of naive  
criteria of local causality, in the form of `advanced effects' 
 in  tree-level scattering processes,  
should not come as a surprise
 (\cf \cite{teleo, lagb}). 

On a more technical level, these theories have no straightforward
Hamiltonian quantization (see however \cite{quim})
 and are  defined, in an operational
sense, via the Feynman diagram expansion \cite{tek, pertbunch}.
 It is thus necessary to
check explicitly the unitarity of the theory. Indeed, in a theory
with built-in nonlocality in time, it is hard to imagine an appropriate
notion of causality which is at the same time useful and nontrivial. It is 
more likely that these theories must be interpreted as S-matrix theories,
in the same sense as critical string theory. Thus, from this very
fundamental point of view, we may regard the existence of a consistent
S-matrix as the weakest possible notion of causality.  

More specifically, it was shown in \cite{ncos, us} that the stringy
regularization of such systems is ill-defined. Following 
\cite{cds, SW}, we can recover certain noncommutative field theories
  (NCFTs) as an appropriate  low-energy limit
of D-brane dynamics in the  background of a constant electromagnetic field.
While this limit is smooth in string perturbation theory for the case of
a background magnetic field, it is problematic for a background electric
field\footnote{The marginal case of a null electromagnetic
background ${\bf E} \cdot {\bf B} = {\bf E}^2 - {\bf B}^2 =0$ is also
smooth, \cf \cite{lightlike}.}.
 This is tied to the well-known instabilities of open-string
dynamics in the presence of electric fields \cite{critf},
 \ie beyond a maximal value
of the electric field, the D-brane becomes effectively tachyonic. Since
the time-noncommutativity is directly related to the electric field,
and this must be large (in string units) in the low-energy limit, one
finds that the low-energy limit always lies on the tachyonic regime
of D-brane dynamics. Therefore, consistency of the string background
requires the scale of time noncommutativity to be of the same order
of magnitude as the string scale.   

The breakdown of the stringy regularization for time-NCFT is a strong
hint at the inconsistency of these models. Still, one can imagine
some contrived analytic continuation of the background parameters,
so that the open-string perturbation theory does converge in the
formal low-energy limit to the series of Feynman diagrams of the time-NCFT.
In particular, a formal continuation of the closed string coupling
$g_s \rightarrow ig_s$ and an exchange of the roles of space and time
in the plane of the electric field would do the job (\cf \cite{us}).  

Therefore, it is desirable to find an internal inconsistency of the
Feynman expansion in time-NCFTs. Such an inconsistency was found
in \cite{jaume}, where a violation of the unitarity cutting rules was
reported in a number of examples. The authors of \cite{jaume} showed
that  scattering amplitudes have extra singularities that cannot be
understood in terms of unitarity thresholds. In itself,
 this {\it does not} 
    furnish
 a no-go argument for time-NCFT, since Feynman diagrams in ordinary
theories are known to present the so-called `anomalous thresholds'
whose interpretation in terms of unitarity rules is very problematic
(see \cite{anal} for a review).

Thus, it is crucial to interpret physically the new 
singularities in order to evaluate the viability of time-NCFTs. One
interesting possibility would be that NCFTs mimic open-string theory
in the sense that unitarity of the S-matrix restricted to open-string
initial and final states {\it requires} the introduction of closed
strings as states in the asymptotic Hilbert space, since they contribute
singularities in intermediate channels.  In an analogous fashion, it
is possible that the S-matrix of time-NCFT becomes unitary once we
add appropriate new states to the asymptotic Hilbert space. 

In fact, such a possibility is hinted at by the simplest example of
a `noncommutative singularity', \ie the case of the normal-ordering
  correction
to the propagator of massless $\phi_*^{4}$ theory in four dimensions. In
the commutative theory (or at the level of planar 
 diagrams in the noncommutative theory) the normal-ordering
 diagram has no analytic
structure, since it contributes a             quadratically divergent 
constant renormalizing the mass. On the other hand, the  
nonplanar diagram, viewed as a one-to-one  scattering amplitude,  is given
by   
\be\label{nice} 
i\CA (p) = -i\lambda \cdot {1\over 6}
\int {d^4 q \over (2\pi)^4}\, e^{i{\tilde p}\,q} \,
{i\over
q^2 +i0} = -i {\lambda \over 24\pi^2} {1 \over p\circ p +i0} 
,\ee
where ${\tilde p}^\mu \equiv p_\nu \theta^{\nu\mu}$   
and  $p\circ p \equiv -{\tilde p}^2$. Thus, we find a contribution to
the imaginary part:
\be
2\,{\rm Im}(\CA) = {\lambda \over 12\pi} \,\delta(p\circ p)
.\ee
Let us now suppose the initial energy positive $p^0 >0$ and a  single  
noncommutative plane  $[x^0, x^1 ] = i\theta_e $.  
 Then we can
write the analog of the optical theorem for this quantity by introducing
$1=\int d^4 k \,\delta(k-p)$ and obtain:
\be
2\,{\rm Im}(\CA) = \int {d^3 {\bf k} \over 2(2\pi)^3 |k_1|} 
\,\left({(2\pi)^2 \lambda \over 6 \theta_e^2}\right) \;\delta(p-k)
.\ee
Thus, we see that the cutting rules can be formally recovered if we introduce
new  states $|\chi_k \ket$ in the asymptotic 
Hilbert space with dispersion relation
$k_0 = |k_1|$. They mix with  the  off-shell 
$\phi$ quanta with
an effective coupling             
$\lambda_{\phi \chi} = \sqrt{4\pi^2 \lambda / 6 \theta_e^2}
$.
All this is strongly reminiscent of the situation one finds 
in nonplanar open-string scattering  amplitudes. In that case, nonplanar 
 amplitudes show new poles,
 without a clear interpretation in terms of open-string
intermediate states. It turns out that they
 just represent the amplitude for an open string
to mix with closed strings. Therefore, in the particular case of the diagram
considered here, the $\chi$ particles are analogous to closed strings. This
analogy is sharpened by considering the `dipole' picture of noncommutative
particles \cite{dipoles}, since this is equivalent to {\it rigid} open
strings. Then, the new states arise from `cutting the dipole' in the
intermediate loop.  

Notice that the effective coupling for producing the $\chi$ particles
 blows-up
in the limit $\theta_e \rightarrow 0$, reflecting a non-analiticity in the
commutative limit --a typical UV/IR mixing. Another peculiar property
of the $\chi$ particles is their lack of propagation in commutative spatial
directions.
The $\chi$ particles could be regarded as excitations of `$\chi$-fields', 
 a generalization of those proposed in \cite{seib, seibb}
in order to reconcile the Wilsonian interpretation of the renormalization
group and the IR divergences found in spatially noncommutative theories.  
The main difference is that these fields are associated to propagating
particles rather than being Lagrange multipliers. Thus, they cannot be
simply considered  as formal devices but {\it must} be included in
any attemp to consistently construct the time-NCFT.

In the following section, we study the general structure of the
`noncommutative singularities' at one-loop and attemp to give an
interpretation along these lines. 
Using results from \cite{windings},
we will show that the unitarity-violating singularities in time-NCFT can be
manipulated into a form that is strongly reminiscent of undecoupled
closed-string modes. Thus, the cutting rules can be formally satisfied
by adding an appropriate set extra  asymptotic degrees of freedom. On
the other hand, unitarity is not restored in a strict sense, because
the extra states are necessarily tachyonic. Thus, time-NCFT appears
to be perturbatively inconsistent, even if we try to add new degrees
of freedom into the problem. Our discussion closes the
conceptual gap  between the string-theorerical arguments of \cite{ncos,us}
and the field-theoretical arguments of \cite{jaume}.

In the last section of the paper we come back to the issue of
unitarity versus causality. We study  in detail some criteria for
microscopic causality at the quantum level. These criteria are
relevant in situations where some Lorentz subgroup 
involving boosts remains unbroken.  In particular, we will expose 
 the impact of the quantum UV/IR mixing on the locality properties
of the theory.

\subsection*{Noncommutative Singularities at One-loop Order} 

\noindent 

For a general one-loop diagram  with $N$ vertices 
in $d$-dimensional  $\phi_*^{n}$ theory we have
\be
i\CA_{\{p_a\}} = {(-i\lambda)^{N} \over |\Gamma|}
 \,\int {d^d q \over (2\pi)^d}\, [{\rm Moyal}] \;\prod_{a=1}^N {i \over
(q+Q_a)^2 - m^2 +i0}
,\ee
where $|\Gamma|$ is the symmetry factor of the diagram, 
and $Q_a +q$ is the momentum running through the $a$'th propagator.
There is an overall momentum conservation delta function $\delta (\sum_a p_a)
$ that we omit in the following. The Moyal phase 
can be factorized into the overall phase of the diagram $\CW_{\rm NC}$,
 depending only
on the external momenta $p_a$, and the nontrivial phase depending
on the loop momentum $q$, given by  
$$
{\rm exp}(i  p_\mu\theta^{\mu\nu}  q_\nu) = {\rm exp}(i{\tilde p}\cdot q) 
,$$
where $p$ is the total momentum flowing through the `nonplanar channel'. 

We can evaluate the loop-momentum integral by introducing Feynman parameters
$x_a$ in the usual fashion:
\ba 
i\CA &=& {\lambda^N (N-1)! \over |\Gamma|} \,\CW_{\rm NC} \,\int 
[dx] \int{d^d q \over (2\pi)^d} {e^{i{\tilde p}\, q} \over
\left[ \sum_a x_a (q+Q_a)^2 -m^2 +i0\right]^N} \nn \\  
& \,& \nn \\
 &=& {\lambda^N (N-1)!\over |\Gamma|} \,\CW_{\rm NC} \, e^{-i{\tilde p}
\cdot \sum x_a Q_a} \,\int [dx] \int {d^d q \over (2\pi)^d} {e^{i{\tilde p}\,q}
\over (q^2 - M_x^2 + i0)^N}
,\ea 
where
$$
\int [dx] \equiv \int_0^1 \prod_a dx_a \,\delta(\sum_a x_a  -1)
,$$
and we have  
 defined an
$x_a$-dependent effective mass
\be
M_x^2 = m^2 - \sum x_a Q_a^2 + (\sum x_a Q_a)^2
.\ee
The resulting momentum integral can be evaluated exactly in terms
of appropriate   Bessel functions (\cf for example \cite{gelfand}):  
\ba\label{bessf}
\CA &=& 2(-\lambda/2)^N \,
(2\pi)^{-{d\over 2}}\; {\CW_{\rm NC} \over |\Gamma|} \nn \\
&\times&  \int [dx]\,
e^{-i{\tilde p} \cdot \sum x_a Q_a} \,\left(M_x^2\right)^{
{d \over 2} -N} \,
{K_{{d\over 2} -N} \left[\sqrt{M_x^2
\, (p\circ p  +i0)}\,\right] \over
\left[\sqrt{M_x^2\,(p\circ p   +i0)}\,\right]^{{d\over 2}-N}}
,\ea
with the convention that  all 
branch cuts are drawn along the negative real axis.  

 The most significant property of this representation
is the occurrence of a 
generic branch-point singularity at  $p\circ p =0$.
The noncommutativity matrix $\theta^{\mu\nu}$  can be invariantly
characterized as space-like or `magnetic', light-like or `null' and  
time-like or `electric' \cite{lightlike}.  In the first two cases, we have
$p\circ p \geq 0$ for real momenta, and  we can only access the
singularity at $p\circ p=0$, \ie the so-called UV/IR singularities
of \cite{seib}.     

On the other hand, in the `electric' case we can access the full
branch cuts along  $p\circ p \leq 0$ with real momenta in the physical region, 
\ie these singularities  resemble particle-production cuts that are
{\it characteristic} of time-NCFT. 
However, the examples studied in \cite{jaume} show that the  
 singularities at $p\circ p <0$  do not satisfy the standard cutting rules, 
\ie they do not have the  standard interpretation
in terms of production of     $\phi$-field quanta. 

Unitarity thresholds in Feynman diagrams are associated to particles
 in a  number
of internal lines going on-shell. 
If a Schwinger parameter $t_a$ is introduced for
each propagator:
$$
{i \over p^2 -m^2 +i0} = \int_0^\infty dt_a\; e^{it_a \,(p^2-m^2+i0)},
$$
 normal thresholds of one-loop diagrams correspond to
exactly two of the Schwinger integrals   being  dominated by the 
region $t_a 
\rightarrow \infty$. Alternatively, we can replace the set of $t_a$ 
parameters by Feynman parameters $x_a$, plus a global Schwinger parameter
$t$, defined by 
$
t=x_a t_a   
$, 
with $x_a \in [0,1]$ and $\sum_a x_a =1$.  
 Then $t=\sum_a t_a$ and {\it any}
cutting of the diagram  produces a singularity associated to the
limit $t\rightarrow \infty$. Therefore, the parametric representation
of (\ref{bessf}) with respect to the $(t, x_a)$ variables is useful
in disentangling the new `noncommutative singularities' from
the usual ones associated with  unitarity cuts.

Taking advantage of the analysis in ref. \cite{windings} we compactify
one spatial direction on a circle of length $L=2\pi R$, which we assume
to be {\it commutative}. We expect this
to produce an effective mass in the nonplanar channel, reminiscent of
massess of closed-string winding modes. This will also allow us to
make more precise the interpretation of ${\tilde p}^2 = -p\circ p $
 as an invariant
mass-squared.

 The effect of the compactification at
the level of the previous diagram is simply to discretize the momenta
in that direction $q_R = n/R$, leading to a measure
\be
 {1\over L} \sum_{n\in {\bf Z}} \int {d^{d-1} q \over (2\pi)^{d-1}} 
.\ee
A convenient way of writing this measure uses the identity 
$
\sum_n \delta(x-n) = \sum_\ell e^{2\pi i \ell x}
$
to rewrite the diagram as 
\be\label{app}
i\CA = {(-i\lambda)^N \over |\Gamma|}
 \,\CW_{\rm NC} \,\sum_{\ell \in {\bf Z}} \int {d^d q \over (2\pi)^d}  
e^{iq\cdot{\tilde p}_\ell } \;\prod_{a=1}^N {i \over (q+Q_a)^2 -m^2 +i0}
,\ee
where ${\tilde p}_\ell$ is a  shifted momentum  given by 
$$
({\tilde p}_\ell)^\mu = p_\nu \theta^{\nu\mu} + L\,\ell\, \delta^\mu_R
.$$
A more technical motivation for introducing the compactification is
apparent in (\ref{app}). Namely if we concentrate on the $\ell \neq 0$ sectors,
the finite size of the circle $L$ acts as an ultraviolet cutoff for
the diagram. This is very convenient, since we would like to disentangle
the occurrence of singularities related to particle production, from those
at $p\circ p =0$, 
inherent to the UV/IR effects of the theory. By selecting the $\ell \neq 0$
sectors we can effectively do so.

We can now introduce Feynman and 
Schwinger parameters as above:
\be\label{param}
i\CA = {(-i\lambda)^N \over |\Gamma|}
 \,\CW_{\rm NC} \,\sum_{\ell \in {\bf Z}} 
\int [dx]\, \int {d^d q \over (2\pi)^d} \,\int_0^\infty dt \,t^{N-1} 
e^{iq\cdot {\tilde p}_\ell   +it\sum_a x_a [(q+Q_a)^2 -m^2 +i0]}
.\ee
Next, we  evaluate the 
  gaussian
integral over $q$ and perform
 a modular transformation of the Schwinger parameter
$s= 1/4t$ to obtain:
\be\label{finala}
i\CA ={(-i\lambda / 4)^N \over |\Gamma|}\,
 {\CW_{\rm NC} \over (i\pi)^{d/2}}
 \,\sum_{\ell } 
\int [dx]\,
e^{-i\sum x_a Q_a \cdot {\tilde p}_\ell} \,\int_0^\infty ds \, s^{-N-1+{d\over
2}}
\,e^{-{i\over 4s} (M_x^2 -i0)} \,e^{-is({\tilde p}^2 - L^2 \ell^2)}
,\ee
where  we have used $({\tilde p}_\ell)^2 = {\tilde p}^2 - L^2 \ell^2$, with
$(L\ell)^2 $ playing the expected
 role of an effective `winding mass', with dimension
length-squared.   This expression may also be obtained directly from  
 (\ref{bessf}) using the integral representation of the Bessel function. 

The singularity structure of this expression is non-standard.
 As stressed before,
normal 
one-loop thresholds must be associated with on-shell intermediate
quanta of the $\phi$ field. However, these contributions come from the
region of integration of   large
proper times $t\rightarrow \infty$ or, equivalently, short dual proper times  
$s\rightarrow 0$. On the other hand, the integral (\ref{finala})
 shows singularity
structure in the opposite limit: $s\rightarrow \infty$,
 which is the ultraviolet
domain in terms of the original $\phi$-field quanta. 

The analytic structure of the amplitudes can be inferred from 
 the representation
(\ref{finala}).  For ordinary theories,
 one defines the amplitude by analytic continuation
from the euclidean region for all momentum invariants (\cf \cite{IZ}).
 In this domain  
$M_x^2 >0$ and the proper-time integral admits a Wick rotation $s\rightarrow
i s$ that makes it convergent at $s=0$. The same Wick rotation renders
the integral convergent in the ultraviolet, $s\rightarrow \infty$,
 provided a convenient cutoff
is in place (in our case, $\ell \neq 0$).  The only novelty in the
 noncommutative  
 case is the requirement of excluding the real branch cut ${\tilde p}^2 \geq
(L\ell)^2$ from  the ordinary euclidean domain, in order to keep the
integral convergent in the $s \rightarrow \infty$ limit. 

This argument shows that any singularity at $-{\tilde p}^2 = p\circ p >0$
is of `ordinary type', since it is  in fact due to the $s\rightarrow 0$ limit.
Thus, `noncommutative singularities' appear as real branch cuts 
for $p\circ p \leq -(L\ell)^2$. Notice that the Wick rotation of the
proper-time parameter $s$ is equivalent    
the regularization of the 
the large-$s$ oscillatory phase by adding a small {\it positive} imaginary
part to the winding mass, \ie $(L\ell)^2 \rightarrow (L\ell)^2 + i0$. Then
 the behaviour in the vicinity of the branch points is given by a series
of terms of the form:  
\be\label{brcut}
\left(p\circ p   +(L\ell)^2 +i0\right)^{k+N-{d\over 2}}\; \left[{\rm log} (
p\circ p   +(L\ell)^2 +i0)\right]
,\ee
where the integer $k$ comes from the Taylor expansion of the phase
containing $M_x^2$, 
and
  the logarithm is present whenever  the number $k+N-d/2$ is a positive integer
or zero. In this formula, all branch cuts are conventionally drawn along
the negative real axis.  

What is the physical interpretation of these branch cuts? Guided by
the example of the normal-ordering diagram in the introduction, together
with the dipole picture,  we would
expect to find a series of pole singularities corresponding to the exchange
of an infinite tower of `winding' $\chi_\ell$ particles of `mass' $|L\ell|$.
These states would naturally descend from closed-string winding modes in
the low-energy limit. 

The general expressions just derived imply that this picture is too naive. 
Namely, the  leading 
$s\rightarrow \infty$ singularity is a pole at $p\circ p = 
-(L\ell)^2$ only for very special diagrams with $N= {d-2 \over 2}$. In
general, it is a softer branch-point singularity, which would suggest
a multi-particle threshold, rather than the single-particle exchange implied
by the dipole picture. 

One possible interpretation of the cut uses a trick 
developed in \cite{seibb},  based on the simple observation that
a branch cut can be viewed as a higher-dimensional pole, \ie 
the negative half-integer powers of $s$ in  (\ref{finala}) may be traded
by a gaussian integral over  
  $d_\perp = 2N+2 -d$
`transverse'  momenta ${\bf z}_\perp$. In this way, we can approximate   
 the amplitude in the vicinity of the $s\rightarrow
\infty$ singularities as:  
\be\label{pole}
\CA \approx \CW_{\rm NC} \;\sum_{\ell\in {\bf Z}} 
 \int
{d^{d_\perp}
 {\bf z}_\perp \over (2\pi)^{d_\perp}} \;{f(p,{\bf z}_\perp, \ell) \over
p\circ p + (L\ell)^2 +{\bf z}_\perp^2 +i0}
,\ee
for an appropriate `coupling' function  $f(p,{\bf z}_\perp, \ell)$ that should
be related to the couplings of $\chi$ particles to the {\it in} and {\it out}
states in the nonplanar channel. This
function has a complicated momentum dependence, which makes this
interpretation rather cumbersome. In addition, the number of extra
`transverse' dimensions $d_\perp = 2N+2 -d$
 is completely {\it ad hoc} and depending on the particular diagram
we consider, unlike the true number of transverse dimensions to a
D-brane.    

It is perhaps more appropriate to interpret the structure
of the singularity in terms of a continuous spectrum of $\chi$ particles,
so that the amplitude is approximated by
\be\label{disprr}
\CA \approx \CW_{\rm NC} \;\sum_{\ell\in {\bf Z}}         
 \int
d\mu^2 
  \;{\rho(p,\ell, \mu^2) \over
p\circ p + (L\ell)^2 +\mu^2 +i0}
,\ee
for an appropriate `spectral density' $\rho(p,\ell, \mu^2)$ that should
be roughly proportional to  the product $\bra out, p, \ell \,|\chi(\mu^2) \ket \bra \chi(\mu^2) \,|
\,in, p, \ell\ket$, and will  have a complicated structure
due to the breakdown of Lorentz invariance. Its explicit form can be
worked out in particular examples from the general expression 
(\ref{bessf}).    

However, even if the spectral density $\rho(\mu^2)$ had the right
properties to be consistent with unitarity, the required 
on-shell condition of the $\chi$ particles,  
$p\circ p + (L\ell)^2 +\mu^2 =0
$, 
would be inconsistent with a positive-energy asymptotic Fock space. 
Solving it  in the simplified situation of  
two orthogonal `electric and magnetic' noncommutative planes:
\be\label{otdis}
\omega_\theta (p) = \sqrt{{\bf p}_e^2 - {1\over \theta_e^2}
\left(\theta^2_m {\bf p}_m^2 +  (L\ell)^2  + \mu^2 \right)}
.\ee
 We see that, in general,  the $\chi$ particles  have 
 tachyonic excitations for nonvanishing  `electric' noncommutativity
$\theta_e \neq 0$. 
 The massless dispersion relation of the single $\chi$
particle in the  example
of the Introduction generalizes   to a continuum of tachyons. In addition,
as soon as $\theta_m \neq 0$,
 the energy squared at fixed `mass' is unbounded from below.    

In our view,  this is the ultimate reason for the breakdown of
unitarity in time-NCFT. Namely, the new singularities really
represent production of tachyonic states, even if the
cutting rules could eventually be satisfied in a formal sense.

\subsubsection*{Connection with String Theory}

\noindent

The  upshot of the discussion in the previous section is that, although
the interpretation of the extra singularities, present for $\theta_e \neq 0$, 
in terms of new `closed-string' particles is very suggestive and even
precise  in some simple
examples, the general structure is rather involved. In addition, it is
found that unitarity is not preserved even if the cutting rules are
formally restored, because the added part of the asymptotic Hilbert
space contains tachyonic states. 

Given the intuitive `dipole' picture and the occurrence of
`winding-like masses' for the effective $\chi$ particles in compact
space, it would be desirable to establish a link between these
{\it ad hoc} degrees of freedom and true closed strings in a model
that would arise from a low-energy limit of string theory. 
The main challenge for such a discussion would be obtaining the
required continuous spectrum of the $\chi$ particles for each
value of the `winding number' $\ell$, as well as the explanation
of the sick dispersion relation  (\ref{otdis}).

Recall that NCFT amplitudes,
when written in terms of Feynman parameters $x_a$ and Schwinger parameter $t$,
descend  directly from the string counterparts in the Seiberg--Witten (SW)
 limit. 
For example, nonplanar annulus amplitudes are integrals over the annulus
modular parameter $\tau$, defined so that the length of the annulus is $2\pi 
\tau$, and Koba--Nielsen parameters $\nu_a \in [0,2\pi\tau)$.  It was 
shown in a series of papers \cite{ampli} that the string  
 amplitude descends in the SW limit to the parametric representation    
(\ref{param}) of the set of low-energy diagrams associated to the given string
diagram. In this process, Koba--Nielsen moduli map to Feynman parameters
according to $\nu_a = 2\pi \tau x_a$ and the annulus modular parameter maps
to the Schwinger parameter via $t=2\pi\alpha' \tau$. 

Under a modular transformation, we can write the string amplitude in the
closed-string channel as an overlap of the closed string propagator between
D-brane boundary states.
\be\label{fac}
\sum_\Psi \left\langle B  |\Psi\right\rangle\left\langle
 \Psi {\Bigg |} \;{g_s^2 \over \sqrt{-{\rm det}(g)}}
\;{-i \over {\alpha' \over 2} \left(g^{AB} p_A p_B - M_{\Psi}^2 
\right) } \,{\Bigg |} \Psi\right\rangle
\left\langle  \Psi  | B' \right\rangle 
,\ee
where $\Psi$ runs over all closed-string  oscillator states, plus the momentum
variables ${\bf p}_\perp$ transverse to the D-brane, and the winding
quantum numbers $\ell$ contributing to   $M_\Psi$: 
$$
M_\Psi^2 = \left({L\ell \over 2\pi\alpha'}\right)^2 + {N_{\rm osc} \over
\alpha'}
.$$
In (\ref{fac}), the dependence of the boundary states on Koba--Nielsen
moduli has been obviated.   The factor of $g_s^2$ comes
from the two boundaries of the cylinder and the determinant factor comes
from the canonical normalization of closed-string states propagating in
a ten-dimensional 
bulk with metric $g_{AB}$. We choose this metric to be unity in the
space transverse to the D-brane. On the other hand, both $g_s$ and
 the world-volume
components of the closed-string metric are determined by formulas in  
  \cite{SW}:
\be\label{clm}
g^{\mu\nu} = \eta^{\mu\nu} - {1\over (2\pi\alpha')^2} \,\theta^{\mu\alpha}
\,\eta_{\alpha\beta} \,\theta^{\beta\nu}
, \qquad 
{g_s^2 \over \sqrt{-{\rm det}(g)}} = G_s^2
,\ee 
where $G_s$ is the effective string coupling, which
in turn determines the NCFT coupling in the SW limit.

In a  
proper-time representation, we write     
$$
{-i \over {\alpha' \over 2} \left(g^{AB} p_A p_B - M_{\Psi}^2 
\right) } = 
\int_0^\infty d\tau_2 \;e^{-i\tau_2 \,\Delta_\Psi}  
,$$
where 
the modular parameter of the cylinder is related to that of the
annulus by $\tau_2 = \pi/\tau$.  
 On the other hand, the dual proper time for the
$\chi$ particles is $s=1/4t = \tau_2 /8\pi^2 \alpha'$. Therefore, the SW
 limit, 
 involving $\alpha' \rightarrow 0$ at fixed $s$,  takes $\tau_2 \rightarrow 0$, 
which {\it is not} the region where the amplitude can be approximated
by a few 
low-lying closed-string states. Rather, in this scaling the whole tower
of closed-string oscillator
 states contributes coherently.
Using the equation (\ref{clm}), the proper-time kernel can be written as:  
\be\label{crit}
\tau_2 \Delta_\Psi  =  
  s \,\left[-p\circ p - (L\ell)^2 +  (2\pi\alpha')^2 \,p^2 - 
(2\pi\alpha')^2 \left(
{\bf p}_\perp^2 + {N_{\rm osc} \over \alpha'} \right)\right]      
,\ee
which is precisely the expression found in the proper-time integral 
 (\ref{finala}): $s({\tilde p}^2 - (L\ell)^2) = -s(p\circ p + (L\ell)^2)$, 
when we send $\alpha' \rightarrow 0$ at fixed $p$ and $\theta^{\mu\nu}$.
The effective gap induced by oscillator states of closed string on the
spectrum of the $\chi$ particles is of order
$$
{{\rm Oscillator \;\;gap} \over |p\circ p|} \sim {\alpha' \over |p\circ p|}
\rightarrow 0
,$$
so that, in the SW limit, the closed-string oscillator spectrum becomes
effectively continuous on the scale of the $\chi$ particles. It follows
that we cannot simply set $\alpha' =0$ in (\ref{crit}) because an infinite
set of modes, labelled by $N_{\rm osc}$ and the transverse momentum to
the D-brane, contribute on the length  scale relevant to the $\chi$ particles.  
This is the interpretation of 
the integral over the  continuous `mass parameter'
$$
\mu^2 \rightarrow 
(2\pi\alpha')^2 \left( {\bf p}_\perp^2 + {N_{\rm osc} \over \alpha'}
\right)
$$
in equation (\ref{disprr}). 
In a  proper treatment one would  write a sum rule  
 for the effective coupling
of the $\chi$ particles, as in \cite{windings},
 involving a trace over all  the  closed-string 
spectrum, and  using
the detailed string amplitude and the known low-energy reduction to
the expression (\ref{finala}).
This trace generates the extra powers of $s$ and the detailed structure
found in (\ref{finala}) that define the properties of the $\chi$ particles,
including the effective transverse dimension $d_\perp =2N+2-d$,
 which {\it is not}
directly related to the  
 true
number of transverse dimensions of the D-brane.  

According to this picture, the  $\chi$ fields are
effective formal devices that represent the coherent coupling of
an infinite number of closed-string states. The fact that they
still resemble ordinary fields in some respects is a nontrivial
property of NCFT, having a degenerate version of the string-theory's
channel duality. 

This construction also explains the tachyonic character of the $\chi$ particles
for $\theta_e \neq 0$.  From the formula (\ref{clm}) for the closed-string
metric, we see that  
$g^{\mu\nu}$ 
only degenerates at
 the NCOS boundary  $|\theta_e| =2\pi\alpha'$, \cite{ncos, us}, which
is the only point where we can argue for the full decoupling of closed
strings. For $|\theta_e| >2\pi\alpha'$, including in particular the
SW limit defining the time-NCFT, 
the role of space and time in the timelike noncommutative plane is
 exchanged \cite{us}.   
 This is exactly what is required to match  the
`empirical' dispersion relations (\ref{otdis})
 introduced above for the $\chi$ particles.  

Therefore, we conclude that the SW limit of strings with timelike
noncommutativity {\it does not} decouple closed-string states, even
in the low-energy limit, since they show up as production of tachyonic  
 states in nonplanar amplitudes. This closes the circle of arguments
in favour of the inconsistency of these theories. In this sense, our
arguments bridge the gap between the criteria of \cite{ncos, us} and
the purely field-theoretical one of \cite{jaume}.

\subsection*{ Tests of Locality  and Causality at the Quantum Level}

\noindent

In ordinary local quantum field theory,
 various technical concepts such as analiticity,
microscopic causality and unitarity are roughly interchangable, due
to some underlying `physical' requirements, such as 
 Lorentz invariance  and   
locality. NCFT violates these two physical conditions in a peculiar
way, keeping still a controlable (and interesting) structure. Thus,
NCFT is a nice laboratory to disentangle the relationships between the
technical criteria cited above. 

In particular, the breakdown of analiticity is to be expected generically in
the case of noncommutative time,
 since the Moyal phases in amplitudes are not analytic functions of
the energy in the upper half plane. This is related to the observed
violations of classical causality criteria in \cite{teleo, lagb}, and also  
 renders invalid most derivations of dispersion relations by means of
Cauchy's theorem (except for the two-point function, since
the global Moyal phase is trivial in this case).  
Thus, we observe a close parallel between the violations of analiticity and
those of unitarity. 

On the other hand, the standard criterion of microcausality (local observables
commute outside the relative light-cone) is more intuitive than the
`technical' criterion based on analiticity, but it is clearly tied to
the underlying Lorentz invariance of the theory. In fact, it loses much
of its significance in situations where the light-cone itself has no
dynamical meaning. Thus, we expect the microcausality criterion to
break down `trivially' already for purely spatial noncommutativity.
Still, since free propagation is Lorentz invariant in NCFT, the breakdown
occurs necessarily  as a result of the interactions and, in view of the
UV/IR effects,  it is an interesting
question to  determine the size of the `causality violation' in perturbation
theory\footnote{For other related  discussions  in different contexts see      
 \cite{caus}.}.

A related interesting question is the following. One can turn on spatial
and/or time noncommutativity while still preserving a certain Lorentz
`little group'. For example, in four dimensions, 
the frame $x^\mu$ satisfying $[x^0,x^1] =i\theta_e, \;[x^2,x^3] =i\theta_m$
preserves a subgroup $SO(1,1) \times SO(2)$ of the four-dimensional Lorentz
group. Thus, boosts along the $x^1$
 axis are still a symmetry even for $\theta_e
\neq 0$, and we  can define a `two-dimensional' light-cone by the
equation $x_e^2 \equiv (x^0)^2 -(x^1)^2 = 0$. In this
 situation, the microcausality
criterion with respect to the four-dimensional light-cone $x^2 = x_e^2 -
x_m^2 =0$ has no particular meaning. However, the same criterion referred
to  the  two-dimensional  light-cone $x_e^2 =0$  {\it is} still
 meaningful.  
             
To analyze the issue we would like to compute the perturbative corrections to
the commutator function
\be\label{commu}
 C(x) = \bra 0 |\,[\phi(x), \phi(0)]\,|0\ket    
.\ee
In fact, it is technically more convenient to
consider the related function given by the difference of retarded and
advanced commutators
\be
{\overline C}(x) \equiv C(x)_R - C(x)_A = {\rm sign} (x^0) \,C(x),
\ee
which in turn is related to the imaginary part of the Feynman propagator 
in position space:
\be
{\overline C}(x) = G_F (x) - G_F (x)^* = 2i\,{\rm Im}\, G_F (x).  
\ee
We can then {\it define} 
microscopic causality by the requirement that ${\rm Im}
\, G_F (x)$ be supported `inside' the light cone, \ie $ {\rm Im} \,G_F (x)
\propto \theta(x^2)$. This
is certainly satisfied at the level of free fields, since the bare propagator
is $\theta$-independent.  The obvious
advantage of this definition for our purposes is that it extends naturally
to the $\theta_e \neq 0$ case, where a Hamiltonian construction of the
commutator from its definition (\ref{commu})
 is absent. In this case, we only have
the Feynman rules as an operational definition of the theory, and one can
readily compute perturbative corrections to ${\rm Im}\, G_F (x)$.

The dressed propagator takes the form
\be\label{propa}
G_F (x) = \int {d^4 p \over (2\pi)^4} \,e^{-ipx} \,{i \over p^2 - m^2 -
\Sigma +i0}
,\ee
where $\Sigma (p_e^2, p_m^2)$ is the 1PI self-energy, a function of
the two $SO(1,1)\times SO(2)$ invariants of the problem:
\be\label{tdm}
p_e^2 = p_0^2 - p_1^2 , \qquad p_m^2 = p_2^2 + p_3^2.
\ee
In terms of these quantities we have
\be
p^2 = p_e^2 - p_m^2 , \qquad p\circ p = \theta_e^2 p_e^2 + \theta_m^2 p_m^2.
\ee
In (\ref{propa}), the definition of $\Sigma$ is such that $m^2$ gives
the `physical' mass after renormalization by the planar diagrams. Thus,
 $\Sigma$ includes the finite part of planar diagrams 
and the contribution from  nonplanar diagrams, which breaks 
 Lorentz invariance. 

On general grounds, we expect the analytic structure of the self-energy
to present the normal thresholds for $p^2 >0$, and the `noncommutative   
thresholds' for $p\circ p <0$. Namely, poles at real positive values of $p^2$
if the theory develops bound states,  the usual multi-particle cuts
for $p^2 \geq 4m^2$, and the noncommutative cuts for $p\circ p <0$ in
the case $\theta_e \neq 0$. 

In what follows,  we will {\it assume} that these singularities in
the real axis exhaust all the singularities of the self-energy function
in the physical sheet as a function of $p_e^2$.   This assumption is
motivated by various considerations. 
 The planar contribution to the self-energy 
is exactly equal to that in the commutative theory, the overall Moyal
phase being trivial due to momentum conservation. In addition, the
nonplanar contributions have a better high-energy behaviour than the
planar ones, and all examples considered show the nonplanar singularities
accumulating in the real $p\circ p \leq 0$ line. Finally, the intuition from
string theory points in the same direction, since ${\tilde p}^2 =
-p\circ p$ is nothing but  the
`Mandelstam variable'  in the closed-string channel.

In performing the Fourier transform to compute $G_F (x)$ it is convenient
to use the corresponding `polar coordinates' with respect to the
$SO(2)$ and $SO(1,1)$ groups.  
 Thus, for the `magnetic' $(x^2, x^3) =x_m$
  plane we change variables from $p_m =(p_2, p_3)$ to the invariant
$p_m^2$ and $SO(2)$ angle, and we can write:  
\be
\int {d^2 p_m \over (2\pi)^2} \,e^{ip_m x_m} \,f(p_m^2) =
{1\over 4\pi} \int_0^\infty dt \,J_0 \left(\sqrt{t x_m^2}\right) \,f(t)
,\ee
for a general function of the magnetic invariant $f(p_m^2)$. In this 
expression, $J_0$ stands for the zeroth-order Bessel function. 

On the other hand, the polar decomposition in the `electric' plane parametrizes
the momenta $p_e = (p_0, p_1)$
 in terms of the invariant $p_e^2$ and the  $SO(1,1)$ rapidity. 
In this case
the contributions to the integral from the different signs of $p_e^2$ must
be considered separately. The complete expression also depends on the
sign of $x_e^2$. Thus, for $x_e^2 >0$ one finds: 
\be
\theta(x_e^2) \,\int {d^2 p_e \over (2\pi)^2} \,e^{-ip_e x_e} \,f(p_e^2) =  
{\theta(x_e^2) \over 4\pi} \,\int_0^\infty ds \,\left[ {2\over \pi}
\, f(-s)\,K_0 \left(\sqrt{s x_e^2}\right) - f(s)\,
 N_0 \left(\sqrt{sx_e^2}\right)\right], 
\ee
where $N_0$ denotes the zeroth-order Neumann function and $K_0$ is the
zeroth-order McDonald function. We may simplify this expression
under the assumption that the otherwise arbitrary function $f(-s)$
admits a `Wick rotation'  $\sqrt{s} \rightarrow -i\sqrt{s}$ in the evaluation
of the integral. Under this
analytic continuation, the Bessel function transforms
$$
K_0 \left(\sqrt{sx_e^2}\right) \rightarrow K_0 \left(-i\sqrt{s x_e^2}\right)
= -{\pi \over 2} \left[ N_0 \left(\sqrt{sx_e^2}\right) -iJ_0 \left(\sqrt{s
x_e^2}\right)\right]
,$$
and the complete integral simplifies, since the $N_0$ functions cancel out.
The final reduction formula is
\be
\theta(x_e^2) \,\int {d^2 p_e \over (2\pi)^2} \,e^{-ip_e x_e} \,f(p_e^2) =
{\theta(x_e^2) \over 4\pi i} \,\int_0^\infty ds \, f(s) \,J_0 \left(\sqrt{
sx_e^2}\right).
\ee
Entirely analogous manipulations give a similar reduction formula for
 $x_e^2 <0$. In this case, the elimination of the Neumann functions
requires the opposite Wick rotation $\sqrt{s} \rightarrow i\sqrt{s}$, so that
$f(s)\rightarrow f(-s)$. The result is
\be
\theta(-x_e^2) \,\int {d^2 p_e \over (2\pi)^2} \,e^{-ip_e x_e} \,f(p_e^2) =
-{\theta(-x_e^2) \over 4\pi i} \,\int_0^\infty ds \, f(-s) \,J_0 \left(\sqrt{
-sx_e^2}\right).
\ee

With these preliminaries, we are ready to write down a general formula
for the modified commutator function, using the general expression for
the dressed propagator in momentum space.

The explicit assumptions   that we need for the analytic structure
of the self-energy can be summarized by demanding that    
\be\label{analp}
{1\over z-t-m^2 -\Sigma(z,t)}
\ee 
has only singularities in the real axis, as a function of the 
complex variable $z$.  These singularities include the usual poles
and cuts for $p^2 >0$, $p\circ p >0$, as well as the `noncommutative
singularities at $p\circ p \leq 0$.  
 In particular,
these conditions
can be explicitly checked for all the  one-loop examples considered
in the literature.   

 There is a component respecting
$SO(1,1)$ microcausality given by:
\be
{\overline C}(x)_{x_e^2 >0} = {i \over 8\pi^2}\,
 \theta(x_e^2)\,\int_0^\infty dt
 \,J_0 \left(\sqrt{tx_m^2}
\right) \,\int_0^\infty ds \;
{\rm Im}\,{J_0 \left(\sqrt{sx_e^2}\right)
\over s-t-m^2 -\Sigma (s,t) +i0} 
.\ee
In addition, the component that violates the $SO(1,1)$ microcausality
is given by
\be\label{viol}
{\overline C}(x)_{x_e^2 <0} = {i \over 8\pi^2}\, 
\theta(-x_e^2)\, \int_0^\infty dt
 \,J_0 \left(\sqrt{tx_m^2}
\right) \,\int_0^\infty ds \; 
{\rm Im}\,{J_0 \left(\sqrt{-sx_e^2}\right) 
\over s+t+m^2 +\Sigma (-s,t) -i0} 
.\ee
 
From these general expressions we see that, as expected, four-dimensional
microcausality is violated as a result of the breakdown of Lorentz invariance.
There is no particular structure depending on $x^2 = x_e^2 - x_m^2$. On the
other hand, there is room for violations of the $SO(1,1)$ microcausality
criterion coming from the term proportional to $\theta(-x_e^2)$.

In general, a non-vanishing imaginary part of
$\Sigma(-s,t)$ in the previous expression implies
a violation of two-dimensional microcausality.
Since $p \circ p=\theta_e^2 p_e^2+\theta_m^2 p_m^2$
in terms of two-dimensional momenta (\ref{tdm}), this 
corresponds,  after the analytic continuation that
leads to (\ref{viol}), to having a contribution to the
imaginary part coming from the cut starting
at $p_e^2=-\theta_m^2 t/\theta_e^2$  and extending
along the negative real axis.  In (\ref{viol}), $s=-p_e^2$.

Another source of $SO(1,1)$-causality violations
is the possibility of having zeroes in the denominator
of the integrand in (\ref{viol}) on the real axis: 
$s+t+m^2+\Sigma(-s,t)=0$.  Viewed as a two-dimensional
dispersion relation this corresponds to the presence
of tachyon poles.

The general structure can be understood by recalling that, under our 
analyticity assumptions, the function in (\ref{analp}) admits a
dispersion relation of the form
\be
{1\over p_e^2 - p_m^2 -\Sigma +i0} = \int_{-\infty}^\infty du  \,
{\sigma(u, p_m^2) \over p_e^2 - u + i0},
\ee
where the `spectral function'
\be
\sigma(u, p_m^2) = ({\rm pole \;\;terms}) \;- {1\over \pi} {{\rm Im}\,
\Sigma(u, p_m^2) \over (u-p_m^2 -{\rm Re}\,\Sigma)^2 + ({\rm Im}\,\Sigma)^2}
\ee
splits naturally in two pieces, $\sigma_{\pm} (\mu^2) = \sigma(\pm \mu^2)$,
respectively  associated to the normal and `noncommutative' 
 thresholds.    

Incidentally,
 we notice that $\sigma_+ (\mu^2)$ has  the interpretation of a honest spectral
function in a K\"allen--Lehmann representation, with positivity ensured
by the unitarity relation ${\rm Im} \,\Sigma \leq 0$ at the
 `normal thresholds'.    
However, a glance at the explicit examples below shows that ${\rm Im}\,\Sigma$
has no definite sign for the `noncommutative thresholds', so that the
spectral interpretation of $\sigma_- (\mu^2)$ is problematic\footnote{
This  particular
case  of (\ref{disprr}) makes it explicit that the spectral
 interpretation of
the function $\rho(p, \ell, \mu^2)$ appearing in that equation
 will  necessarily involve {\it indefinite norm} in
the asymptotic Hilbert space of the $\chi$ particles.}.

Taking the appropriate Fourier transforms we can give a two-dimensional
`spectral representation' of the commutator function ${\overline C}(x) =
{\overline C}_+ (x) + {\overline C}_- (x)$, where:
\be
{\overline C}_\pm (x) = \int_{\mu_\pm^2}^\infty d\mu^2 \,g_\pm (x_m, \mu^2)\,
{\cal C}_e (x_e)_{ \pm \mu^2}, 
\ee
with   
$$
g_\pm (x_m, \mu^2) = \int {d^2 p_m \over (2\pi)^2} \,e^{ip_m x_m} \,
\sigma_\pm (p_m^2, \mu^2) 
$$
and ${\cal C}_e (x_e)_{ \pm \mu^2}$ denoting the two-dimensional {\it free}
 commutator
functions with mass squared $\pm \mu^2$:
\be
{\cal C}_e (x_e)_{\pm \mu^2} = \pm{\theta(\pm x_e^2) \over 2i} \,J_0 \left(
\sqrt{\pm \mu^2 x_e^2}\right).
\ee
Thus, we see that violations of two-dimensional
  microcausality are associated to
the contribution of the `spectral function' $\sigma_-$ in the tachyonic
branch, \ie the noncommutative thresholds for emission of $\chi$ particles.

This shows that our  $SO(1,1)$-invariant causality criterion is equivalent
to unitarity.    
In particular, NCFTs with space-like noncommutativity are `causal' by
the two-dimensional criterion.

\subsubsection*{Some Examples}  

\noindent

It is instructive to check the previous general statements with some
simple examples. We consider the one-loop normal-ordering correction in 
massless $\phi_*^{4}$ theory, with renormalization conditions so that
the `physical' mass vanishes after planar renormalization. Then,
the nonplanar  graph is given by
\be
\Sigma(p_e^2, p_m^2) = i {\lambda \over 6} \int {d^4 k \over (2\pi)^4} 
{e^{ik\cdot {\tilde p}} \over k^2 +i0 } = {\lambda \over 24\pi^2} {1 \over
p\circ p +i0} = {\lambda \over 24\pi^2} {1\over \theta_e^2 p_e^2 + \theta_m^2
p_m^2 +i0}
.\ee
There are no finite contributions to ${\rm Im}\,\Sigma (\pm s, t)$, and the
only source for ${\overline C}(x)$ comes from the pole terms. 
In the case of purely spatial noncommutativity one finds
\be\label{spacel}
{\overline C}(x)_{\theta_e =0} = {1\over 8\pi i} \theta(x_e^2 ) \,\int_0^\infty
dt\,J_0 \left(\sqrt{tx_m^2}\right) \,J_0 \left(\sqrt{x_e^2}\,
 \sqrt{t +g^2/t}\right)
,\ee
where
$
 g^2 \equiv \lambda / 24\pi^2 \theta_m^2
$.

It is interesting to notice the $t\rightarrow g^2 /t$ `duality' in
the argument of the second Bessel function; a rather transparent
manifestation of the    UV/IR effects.  
 One can also check, using identities
of Bessel functions, that  ${\overline C}(x) \propto \delta(x^2)$ for  $g=0$,
\ie the modified commutator at $g^2=0$
 is supported on the four-dimensional light-cone.
 This corresponds
either to the free-field case, $\lambda =0$, 
 or to the case of infinite noncommutativity, $\theta_m = \infty$.  

As expected, the violations of microscopic causality for purely spatial
noncommutativity are tied to the breaking of Lorentz invariance. Thus,
${\overline C}(x)$ is non-zero outside the four-dimensional
 light-cone $x^2 =x_e^2 -x_m^2 \leq 0$. 
On the other hand, the two-dimensional microcausality is not violated, since
${\overline C}(x)$ vanishes for $x_e^2 <0$. It is also interesting to notice
that ${\overline C}(x_m, x_e^2 =0) =0$ for any value of $x_m^2$. 

An asymptotic expansion of 
 (\ref{spacel}) for large $x_m^2$ at constant ratio $a = x_e^2 /x_m^2 <1$
 may be obtained by a saddle-point approximation:   
$$
{\overline C}(x)_{\theta_e =0} \propto {1\over a} \,
{g^{1\over 4} \over |x_m |^{3\over 2}} \sim {1\over a} \,{\lambda^{1\over
8} \over \theta_m^{1\over 4} |x_m|^{3 \over 2} } 
,$$
modulated by an oscillatory phase of frequency $\sqrt{2g a x_m^2}$.  
We see that the non-locality is of long range, presumably as
 a consequece of the
UV/IR mixing.

In the case $\theta_e \neq 0$, the special choice $\theta_e = \theta_m \neq 0$
is convenient to simplify the analysis, while maintaining all 
 the qualitative features 
 intact. The contributions to ${\overline C}(x)$ are again of pole type. 
However, now there is also a term proportional to $\theta (-x_e^2)$:   
\be
{\overline C}(x) = {1\over 8\pi i} \int_0^\infty dt
J_0 \left(\sqrt{tx_m^2}\right) \left[ \theta(x_e^2) {t+m_t^2 \over 
2m_t^2} J_0 \left(m_t \sqrt{x_e^2}\right) + \theta(-x_e^2){t-m_t^2 \over
2m_t^2} J_0 \left(m_t \sqrt{-x_e^2}\right)\right]
\ee   
where 
$
m_t^2 \equiv \sqrt{t^2 +g^2}
$, and $g$ is still given as above. 

If we interpret the pole in two-dimensional terms, the term proportional
to $\theta(-x_e^2)$ comes from a `tachyonic' excitation. 
 One can calculate exactly the integral for $x_m^2 =0$ in
terms of Bessel and Thomson functions. 
The piece that violates $SO(1,1)$ causality yields  
\be
\theta(-x_e^2)\,{\overline C}(x)_{\theta_e = \theta_m} = {i\over 8\pi}\,
g\,\left[ {J_1 (\xi \sqrt{2}) \over \xi \sqrt{2}} -{\rm kei}_1 (\xi)\,
{\rm bei}_1 (\xi) + {\rm ber}_1 (\xi) \,{\rm ker}_1 (\xi)\right],
\ee
where $\xi \equiv \sqrt{-g x_e^2 /2}$.   
The leading
 asymptotic behaviour for $x_m^2 =0$ and large $\sqrt{-x_e^2}$ is
$$
{\overline C}(x)_{\theta_e = \theta_m} \propto \sqrt{g\over -x_e^2} 
\sim {\lambda^{1\over 4} \over \sqrt{-\theta_e \,x_e^2}},
$$
modulated by an oscillating phase of frequency $\sqrt{-gx_e^2}$, so that
 the violation of two-dimensional microcausality is also
of long range in this case.  

Another instructive case is that of two-dimensional massive scalars. Here,
Lorentz invariance is never broken, but time is always noncommutative.  The
general formula for the commutator function is derived along similar
lines and one obtains
\be
{\overline C}(x) = {i\over 2\pi} \int_0^\infty ds\left[{\rm Im}\,{
\theta(x^2) J_0 \left(\sqrt{sx^2}\right)\over s-m^2 -\Sigma(s) +i0} 
+ {\rm Im}\, {\theta(-x^2) J_0\left(\sqrt{-sx^2}\right) \over s+m^2 +
\Sigma(-s) -i0}\right]
.\ee
Taking for the self-energy the nonplanar normal-ordering graph  one finds
\be
\Sigma(s) = 
 {\lambda   \over 12\pi} \,K_0 \left(m\theta 
\sqrt{s +i0}\right)
.\ee 
In fact, in this case one may evaluate the commutator in a power series
in $\lambda$, since the corresponding integrals are convergent. The leading
correction outside the light-cone is given by the function  
\be
{\overline C}(x)_{x^2 <0}  = {i\lambda \over 24\pi} \left\{ \begin{array}{cl}
{\sqrt{-x^2} \over 2m} \,I_1 (m\sqrt{-x^2}) \,K_0 (m^2 \theta) -\theta \,
I_0 (m\sqrt{-x^2}) \,K_1 (m^2 \theta), & \mbox{if $\sqrt{-x^2} < m\theta$} \\
& \\
\theta \,
K_0 (m\sqrt{-x^2}) \,I_1 (m^2 \theta) -
 {\sqrt{-x^2} \over 2m} \,K_1 (m\sqrt{-x^2}) \,I_0 (m^2 \theta), 
& \mbox{if $\sqrt{-x^2} > m\theta$}   
\end{array} \right.
\ee
 This function increases away from the
light cone to reach a maximum  around $\sqrt{-x^2} \approx m\theta$. Then it
decreases with exponential asymptotics ${\rm exp}(-m\sqrt{-x^2})$
 controlled just
by the mass of the field. Therefore, we have the expected behaviour,
with breakdown of microcausality in spite of the preservation of
Lorentz invariance. In this case however the light-cone fuzziness
at ${\cal O}(\lambda)$ 
is of short range, with a width of order $m \theta$.   

For completeness, we quote here the result of the same calculation
for the light-like case \cite{lightlike},  
 with $\theta^{02} = \theta^{12} \equiv \theta$, so that $p\circ p =
\theta^2 (p_0 - p_1)^2 = \theta^2 p_-^2$.  
Defining $\alpha
 \equiv g^2 /(x^+)^2$ and $ \beta \equiv -x^2 /12$, we find the exact result 
\be
{\overline C}(x)_{\rm lightlike} = {i \over 8\pi^2} \int_0^\infty
du \,{\rm cos} \left({x^2 \,u \over 4} + {\alpha \over u^3}\right)
=-{8^{1/3}\sqrt{-g^2 x^2} \over 24 |x^+|}\,  G^{30}_{04}
\left[\alpha^2\beta^3;\shalf,
\ssixth,-\ssixth;-\shalf\right]
,\ee
where the second expression in terms of the
 Meijer G-function assumes $x^2 <0$. Outside the four-dimensional
light-cone, for large negative values of $x^2$, this function
decays exponetially as ${\rm exp}\left(-(\sqrt{-x^2} / \ell)^{3/2}\right)$,  
 with a characteristic length $\ell = (\theta x^+ /\sqrt{\lambda})^{1/3}$.  
Notice, however, that the light-like configuration leaves no remnant of the
four-dimensional light-cone. Therefore, this violation of four-dimensional
microcausality is expected because of the breakdown of Lorentz invariance.
It is interesting to note that the detailed form of the causality violation
at fixed $x^+$ is of short range, \ie exponentially supressed.

\subsection*{Conclusions}

\noindent

In this note 
we  reexamine various  issues of unitarity and causality
 in perturbative NCFT with time/space
noncommutativity.    
 We discuss   general one-loop diagrams and we   
 establish the locus of their  `noncommutative singularities' 
 to be
$p\circ p \leq 0$. `Noncommutative singularities' refers to those singularities
that have no simple interpretation in terms of standard cutting rules. The
first examples of these singularities were found in \cite{jaume}.

 More generally, if the  theory is compactified
on a circle of {\it commutative} length $L$, there are  infinite 
 sectors,   
labelled by an integer
 `winding number $\ell$', with noncommutative singularities
 for $p\circ p \leq -(L\ell)^2$.    
These results are robust in the sense that they follow from analytic
properties of parametric representations
and do not rely on combined analytic continuations in momenta {\it and}
the noncommutativity matrix $\theta^{\mu\nu}$, as in \cite{jaume}. 

We can interpret these singularities in terms of new asymptotic states, the
$\chi$ particles, 
analogous to closed strings.
 One considers an asymptotic Hilbert space of the form
$
{\cal H}_{\infty} = {\cal H}_\phi \otimes {\cal H}_\chi$, where ${\cal H}_\phi$
is the free Fock space of the $\phi$-field quanta and ${\cal H}_\chi$
is the Fock space of $\chi$ particles.  Then, a formal restoration of  the  
cutting rules requires the $\chi$ particles to have a continuous spectrum
 and, in general, a tachyonic dispersion relation.  Morever, a closer look
at explicit examples for the two-point function reveals that ${\cal H}_\chi$
necessarily has {\it indefinite norm}. 

Therefore, time-NCFT is perturbatively inconsistent because 
unphysical  excitations, \ie `ghostly tachyons', are produced in scattering.
 This is similar to the
breakdown of string perturbation theory beyond the NCOS barrier
\cite{ncos, us}. In models with a string regularization, we 
 specifically  show that 
  the $\chi$ particles
descend from closed strings in the low-energy limit.  This  provides a
microscopic understanding of the peculiar properties of the $\chi$ particles,
namely the continuous spectrum, the `winding masses'
 and the tachyonic dispersion relation. It also unifies the field-theoretical
and stringy  no-go arguments against time-NCFT.

We have also carried out a quantitative study of some causality criteria.
Since NCFTs are theories of extendend `dipoles', local microcausality
defined in terms of commutators of local observables, such as
$[\phi (x), \phi(x')]$, loses much of its meaning, \ie the `light-cone'
has no significance due to the breaking of Lorentz invariance. However,
if $\theta^{\mu\nu}$ is block-diagonal  with a two-dimensional eigenspace
including time, there is an unbroken subgroup of the Lorentz group including
$SO(1,1)$ boosts, and a weaker notion of microcausality can be defined  
with respect to the corresponding two-dimensional light-cone,  independently
of whether the time coordinate is commutative or noncommutative.  

We compute quantum corrections to the commutator of scalar fields and
find that $SO(1,1)$ microcausality is broken if and only if unitarity is
broken. This result lends support to the intuitive idea that space-time
nonlocal theories must be interpreted as S-matrix theories. Thus `causality'
criteria must be referred  to consistency conditions of the S-matrix.

\subsubsection*{Acknowlegdments} 

\noindent

L.A-G. would like to thank the hospitality of the
 Physics Department of the Humboldt University
at Berlin,  where part of this work was done, and in particular
 Dr. Dieter L\"ust.
 The work of R.Z. is supported by the Swiss National Science Foundation.

\end{document}